\documentclass[aps,prd,amsmath,superscriptaddress,nofootinbib,twocolumn]{revtex4-1}

\usepackage{varwidth}
\usepackage{multirow}
\usepackage{dcolumn}
\usepackage{tabularx}
\usepackage{booktabs}
\newcolumntype{C}{>{\centering\arraybackslash}X}
\usepackage{graphicx}
\usepackage{epstopdf}
\usepackage{graphics}
\usepackage{epsfig}
\usepackage{epstopdf}
\usepackage{makecell}
\usepackage{bm}
\usepackage{amsmath}
\usepackage{amsfonts}
\usepackage{amssymb}
\usepackage{hyperref}
\usepackage[mathlines]{lineno}

\usepackage{indentfirst}
\usepackage{xcolor}
\usepackage[capitalize]{cleveref}
\usepackage{tabularx}

\usepackage{hyperref}
\usepackage{cuted}
\usepackage{lipsum}
\hypersetup{
    colorlinks=true,
    linkcolor=blue,
    citecolor=blue,
    urlcolor=blue,
    anchorcolor=blue}

\begin{document}

\preprint{}

\title{A comparative study of $T_{cc}$ versus $X(3872)$ production in $pp$ collisions at $\sqrt{s}=$ 7 TeV}

\author{Hongge Xu}%
%\email{xuhg@cug.edu}
\affiliation{Institute of Astronomy and High Energy Physics, School of Physics and Electromechanical Engineering, Hubei University of Education, Wuhan 430205, China}
\author{Tianqi Luo}
\affiliation{College of Physical Science and Technology, Central China Normal University, Wuhan 430079, China}
\author{Yi-Long Xie}
\affiliation{School of Mathematics and Physics, China University of Geosciences, Wuhan 430074, China}
\author{Zhi-Lei She}
\email{shezhilei@cug.edu.cn}
\affiliation{School of Mathematics and Statistics, Wuhan Textile University, Wuhan 430200, China}

\author{Ning Yu}
\email{ning.yuchina@gmail.com}
\affiliation{Institute of Astronomy and High Energy Physics, School of Physics and Electromechanical Engineering, Hubei University of Education, Wuhan 430205, China}

\author{Zuman Zhang}
\affiliation{Institute of Astronomy and High Energy Physics, School of Physics and Electromechanical Engineering, Hubei University of Education, Wuhan 430205, China}
\date{\today}

\begin{abstract}
The production of exotic hadrons $T_{cc}$ and $X(3872)$ in $pp$ collisions at $\sqrt{s}=7$ TeV is compared using the parton and hadron cascade model PACIAE together with the dynamically constrained phase-space coalescence model DCPC. In the simulation, the compact tetraquark state and the loose molecular state are formed in the partonic and hadronic levels, respectively. Our analysis of the transverse momentum spectra reveals a significant discrepancy between the compact state and the molecular states. Furthermore, the production asymmetry between $T_{cc}^+$ and  $T_{cc}^-$ is investigated.
Finally, the coalescence parameters are extracted from the calculated spectra to further characterize the emission source properties.
These distributions are proposed as valuable criteria for distinguishing between these states and investigating their internal structures in experimental measurements.
\end{abstract}
%\begin{description}\item[PACS numbers]
%\verb+25.75.Nq, 24.10.Lx, 24.10.Pa+
%\end{description}

%\pacs{  }
%\keywords{Exotic state, $T_{cc}$, $X(3872)$}
\maketitle

\flushbottom
\section{Introduction} \label{sec1}
The quark model has been an important tool for understanding the low energy region of Quantum Chromodynamics (QCD), and unveiling the internal structure of hadrons~\cite{Gell-Mann:1964ewy,Zweig:1964jf}. The majority of observed hadrons can be classified within this framework as baryons with three quarks or mesons composed of a quark-antiquark pair. However, the requirement of color neutrality alone does not preclude the existence of more complex configurations, such as tetraquarks and pentaquarks, etc. Indeed, the existence of exotic states was also suggested by Gell-Mann~\cite{Gell-Mann:1964ewy} and Zweig~\cite{Zweig:1964jf} at the early stage of the quark model. Such exotic hadrons, which lie beyond the conventional configurations, provide a unique window to decode the mystery of hadronization.

To date, dozens of so-called XYZ exotic states lying outside the conventional quark model are observed in experiments, and various theoretical investigations have been proposed for these new forms of matter, such as compact tetraquarks~\cite{Esposito:2014rxa,Lebed:2016hpi}, hadronic molecules~\cite{Guo:2017jvc}, hybrids~\cite{Meyer:2015eta}, normal heavy quarkonium~\cite{QuarkoniumWorkingGroup:2004kpm}, and so on, with the potential to disclose new information about the fundamental strong force.
Despite the abundance of theoretical interpretations regarding the properties of exotic states, controversial interpretations are often seen in the literature. To deepen our understanding of exotic hadrons, physicists are putting tremendous effort into exploring novel research methods and topics.

The first candidate of the exotic state, $X(3872)$ with
quark content of $c\bar{c}u\bar{u}$ or $c\bar{c}d\bar{d}$, was observed by Belle Collaboration in 2003~\cite{Belle:2003nnu}.
%Since then, the study of exotic hadrons has entered a new era, with numerous multiquark candidates being observed experimently, such as XYZ tetraquarks and $P_c$ pentaquarks, and so on ~\cite{Yamaguchi:2019vea,Brambilla:2019esw,Chen:2022asf,ParticleDataGroup:2024cfk}.
Recently, the LHCb Collaboration announced the observation of the first doubly charm tetraquark, $T_{cc}(3875)$ with quark content of $cc\bar{u}\bar{d}$ or $\bar{c}\bar{c}ud$, near the $DD^*$ threshold~\cite{LHCb:2021auc,LHCb:2021vvq}.
%As it does not mix with conventional mesons, and it is regarded as an ideal candidate for an exotic hadron.
The masses of $T_{cc}(3875)$(open-charm state) and $X(3872)$(hidden-charm state) are very close. Their striking characteristics have sparked much discussion about their nature, such as the observations in experiments~\cite{LHCb:2024tpv,BESIII:2023hml,LHCb:2020xds} and theoretical interpretations~\cite{Xu:2023lll,Esposito:2021vhu,Brambilla:2024imu,Hua:2023zpa} for $T_{cc}$ and $X(3872)$ indicate that they could be a compact tetraquark or a loosely bound molecular configuration composed of charmed mesons.
Therefore, in this work, we provide further insight and perform a comparative study of the production of $T_{cc}$ and $X(3872)$.

%Phenomenological models remain primary tools for unveiling the nature of experimentally observed exotic candidates.
 In this paper, we explore the production of $T_{cc}$ (including $T_{cc}^+$ and its antimatter counterpart $T_{cc}^-$) and $X(3872)$ in $pp$ collisions at $\sqrt{s} =7$ TeV employing the parton and hadron cascade model (PACIAE)~\cite{Lei:2023srp} and the dynamically constrained phase-space coalescence model (DCPC)~\cite{Yan:2011fq}.
 Both compact tetraquark and hadronic molecular frameworks are considered, which can both describe their mass positions but involve different underlying physics. Firstly, the production asymmetry between $T_{cc}^+$ and $T_{cc}^-$ is investigated.
 To investigate the physical nature of $T_{cc}$ and $X(3872)$, we then compare their yields and kinematic distributions (i.e., transverse momentum $p_T$ and coalescence parameters).

The remainder of this paper is organized as follows: The methodologies are detailed in Section: Method~\ref{sec:method}. Section: Results and discussion~\ref{sec3} presents the numerical results and analysis. Finally, our findings and conclusions are discussed in the last Section~\ref{sec4}.

\section{method} \label{sec:method}
\subsection{PACIAE model}
The PACIAE model~\cite{Sa:2011ye,Lei:2023srp} is a phenomenological model based on the PYTHIA code~\cite{Sjostrand:2006za}, which can be used successfully to simulate relativistic elementary particle collisions and nucleus-nucleus collisions~\cite{Sa:2011ye,Lei:2023srp,Sa:2013csa,Zheng:2018yxq}.
With the PACIAE model framework, a $pp$ collision is divided into the initial parton stage, the parton rescattering stage, the hadronization stage, and the hadron rescattering stage.

In the first stage, the collision of the proton-proton pair is executed by PYTHIA~\cite{Sjostrand:2006za} with hadronization temporarily turned-off. Consequently, an initial partonic state is established following the QCD hard scattering and the associated initial- and final- state radiations. Here a process of the gluons breaking-up and energetic quarks and antiquarks deexcitation is then executed~\cite{Lei:2023srp}.
In the subsequent stage, the lowest-order perturbative quantum chromodynamics (LO-pQCD) parton-parton interaction cross sections~\cite{Combridge:1977dm} are used. The results of partonic final state (PFS) comprising numerous quarks and antiquarks with four-coordinates and four momenta, are generated after partonic rescattering.
During the hadronization state, the Lund string fragmentation regime or the coalescence model is implemented to generate an intermediate hadronic state, and the latter is chosen in this work.
The final stage is hadronic rescattering, which produces the hadronic final state (HFS), which consists of abundant hadrons with four coordinates and four momenta. %The schematic structure of above transport processes is shown in Fig. 1.

\subsection{DCPC model}
The dynamically constrained phase-space coalescence (DCPC) model was originally proposed to investigate the production of light nuclei and anti-nuclei in $pp$ and nucleus-nucleus collisions~\cite{Yan:2011fq,She:2020qyp} and has since been successfully extended to calculate the production of exotic hadrons including the $X(3872)$, $Z_c(3900)$, $P_c$, following simulations with the transport model PACIAE~\cite{PhysRevC.110.014910,Wu:2023aim,Xu:2021drf,Chen:2021ifb,Zhang:2020vfv}. The DCPC model can be used to form the compact tetraquark state and the loose molecular state in the partonic and hadronic levels, respectively.

Grounded in the principles of quantum statistical mechanics~\cite{Stowe_2007,Kubo1965}, the yield of an N-particle cluster (formed by partons in PFS or hadrons in HFS) is defined as :
\begin{equation} \label{eq1}
Y_N=\int_{E_\alpha\leqslant E\leqslant E_\beta}\prod_{i=1}^{N}\frac{d\vec{q}_id\vec{p}_i}{h^{3N}},
\end{equation}
where $E_\alpha$ and $E_\beta$ denote the lower and upper energy thresholds of the cluster, respectively. $\vec{q}_i$ and $\vec{p}_i$ represent the three-dimensional coordinate and momentum of the $i$th particle. For a valid cluster to form naturally, the constituent particles should satisfy three certain constraints, component constraint, coordinate constraint, and momentum constraint.

For the tetraquark state, it is coalesced (hadronized) in PFS using the DCPC model by four component quarks, specifically, charm quark $(c/\bar{c})$ and light quark $q(\bar{q})$ (with $q = u/d$, the same notation applies hereafter). The yield of the tetraquark state is calculated by
\begin{equation} \label{eq2}
    Y_{\text{Tetraquark}}=\int_{E_\alpha\leqslant E\leqslant E_\beta}{\delta_{1234}\frac{\prod_{i=1}^{i=4}d\vec{q}_id\vec{p}_i}{h^{12}}},
\end{equation}
where
\begin{equation} \label{eq3}
\delta_{1234} =
\begin{cases}
1 & \text{if\quad }
\begin{array}{l}
1\equiv c/\bar{c},2\equiv c/\bar{c},3\equiv u/\bar{u}/d/\bar{d},4\equiv u/\bar{u}/d/\bar{d}, \\
2~m_{D}\leqslant m_{\text{inv}} \leqslant 2~m_{D^{*}}, \\
|\vec{q}_{i0}| \leqslant R_0\ (i=1,2,3,4),
\end{array} \\
0 & \text{otherwise.}
\end{cases}
\end{equation}

In Eq.~(\ref{eq3}), $R_0$ and $|\vec{q} _{i0}|$ denote the radius of the tetraquark state and the relative distance between the $i$th constituent quark (or antiquark) and the cluster's center of mass, respectively. The $m_{\text{inv}}$ is the invariant mass of the four-component system, calculated as:
 \begin{equation}  \label{eq4}
m_{\text{inv}}=\sqrt{{\left(\sum_{i = 1}^{4}E_i\right)}^2-{\left(\sum_{i = 1}^{4}\vec{p_i}\right)}^2},
 \end{equation}
where, $E_i$ and $\vec{p}_i$ ($i=1,2,3,4$) are the energy and three-momentum of the $i$th constituent quark.

Similarly, the yield of the molecular state is defined as
\begin{equation} \label{eq5}
  Y_{\text{Molecular}}=\int_{E_\alpha\leqslant E\leqslant E_\beta}{\delta_{12}\frac{d\vec{q}_1d\vec{p}_1d\vec{q}_2d\vec{p}_2}{h^{6}}.}
\end{equation}
In the above equation, $\delta_{12}$ is expressed as

\begin{equation}
\label{eq6}
\delta_{12} =
\begin{cases}
1 & \text{if\quad }
\begin{array}{l}
1\equiv D,2\equiv D^{*}, \\
2~m_{D}\leqslant m_\text{inv}\leqslant 2~m_{D^*}, \\
|\vec{q}_{i0}| \leqslant R_0(i = 1,2),
\end{array} \\
0 & \text{otherwise.}
\end{cases}
\end{equation}
where $|\vec{q}_{i0}|$ is the relative distance between the $i$th component meson ($D$ or $D^*$ meson in this work) and the molecular cluster's center of mass. The $R_0$ is the radius of the molecular state. The invariant mass $m_{\text{inv}}$ is calculated as $m_\text{inv}=\sqrt{{\left(E_1+E_2\right)}^2-{\left(\vec{p}_1+\vec{p}_2\right)}^2}$, where $E_1$, $E_2$ and $\vec{p}_1$, $\vec{p}_2$ are the energy and three-momentum of the component particle $D$ mesons, respectively.

\section{Results and Discussion} \label{sec3}
\subsection{Production and yield}
In the paper, we utilize PACIAE 3.0 and DCPC model to simulate the production of $T_{cc}^+$, $T_{cc}^-$ and $X(3872)$ in $pp$ collisions at $\sqrt{s}=7$ TeV.
In our framework, the production of exotics involves a two-step process: the generation of their constituents and the subsequent formation of the exotic state. The primary PACIAE model parameters of the allowed number of deexcitation generation and the threshold energy of deexcitation the energetic (anti)quark deexcitation process, i.e., adj1(16)=1 and adj1(17)=1.8 GeV, are fixed by fitting the ALICE data of $D^0$ and $D^+$ yield in $pp$ collisions at $\sqrt{s}=2.76$ TeV~\cite{ALICE:2012inj}.
Subsequently, the tetraquark state and molecular state are formed in PFS and HFS by DCPC model, respectively.

In tetraquark picture, we first generated component quarks in PFS by PACIAE. The compact tetraquark state is then coalesced (hadronized) via four component particles of $c$($\bar{c}$) and $q$ (anti)quarks ($q=u/d$) by DCPC model. The components of $T_{cc}^+$, $T_{cc}^-$ and $X(3872)$ are $cc\bar{u}\bar{d}$, $\bar{c}\bar{c}ud$ and $c\bar{c}u\bar{u}$, respectively.
Taking into account the phase space constraints in DCPC model, $R_0<1.0$~fm are set, treating the compact state as a pointlike particle\cite{Chen:2021akx,Grinstein:2024rcu}, and set to $3729.68<m_\text{inv}<4020.52$ MeV for $T_{cc}^+$ and $T_{cc}^-$, $3729.68<m_\text{inv}<4013.70$ for $X(3872)$\cite{Zhang:2020dwn}.

Similarly, in the molecular picture, they are hadronized
by the coalescence of two mesons in the FHS. In the paper, we simulate the production of the $D^0D^{*+}$ and $D^+D^{*0}$,  $D^0D^{*-}$
and $D^-D^{*0}$, $D^0\bar{D}^{*0}$ and $D^{*0}\bar{D}^0$ for
the formation of the $T_{cc}^+$, $T_{cc}^-$ and $X(3872)$, respectively.
The constraints for Eq.~(\ref{eq6}) are set as follows: $3729.68<m_\text{inv}<4020.52$ MeV for $T_{cc}^\pm $, $3729.68<m_\text{inv}<4013.70$
for $X(3872)$\cite{Zhang:2020dwn}, and 1.0 fm $< R_0 <$ 10.0 fm
for the molecular state ($R_0$ should be larger than the sum of
the radius of two component mesons and less than
the interaction range of 20 fm).

In total, we simulate 600 million $pp$ collision events to
generate molecular states, and 18.6 million $pp$ collision events
for tetraquark states. The yield is predicted and summarized
in Table \ref{table1}.

\begin{widetext}
\centering
\begin{table*}[!htpb]
\caption{The PACIAE+DCPC model simulated yield of the molecular and tetraquark $T_{cc}^+$, $T_{cc}^-$ and $X(3872)$ in $pp$ collisions at $\sqrt{s}=7$ TeV, without any kinematic constraint. The uncertainties are the statistic ones.}
\renewcommand{\arraystretch}{1.5}
\label{table1}
\begin{tabular}{cccc} \hline \hline
   &{$T_{cc}^+$} &$T_{cc}^-$  & {$X(3872)$}        \\ \hline
Molecular     & $(7.55\pm 0.347)\times 10^{-7}$  & $(7.55\pm 0.358)\times 10^{-7}$ & $(5.91\pm 0.100)\times 10^{-6}$  \\
Tetraqurak    & $(2.94\pm 0.126)\times 10^{-5}$  & $(2.90\pm 0.125)\times 10^{-5}$ & $(11.09\pm 0.077)\times 10^{-4}$ \\
Ratio(T/M)    & $38.88\pm 2.470$  & $38.44\pm 2.460$ & $187.55 \pm 3.432$ \\
\hline \hline
\end{tabular}
\end{table*}
\end{widetext}

As shown in Table~\ref{table1}, the molecular state is more likely to be produced than the tetraquark state. Meanwhile, the yield of the $X(3872)$ state is significantly higher than those of $T_{cc}^+$ and $T_{cc}^-$ for both the molecular and tetraquark configurations. This disparity is attributed to the fact that the yield of $c\bar{c}$ pairs is statistically larger than that of $cc$ or $\bar{c}\bar{c}$ pairs. Consequently, $X(3872)$ is naturally more abundant than double-charm states $T_{cc}$.
\subsection{Transverse momentum distribution}
\begin{figure}[htbp!]
  \centering
    \includegraphics[width=0.5\textwidth]{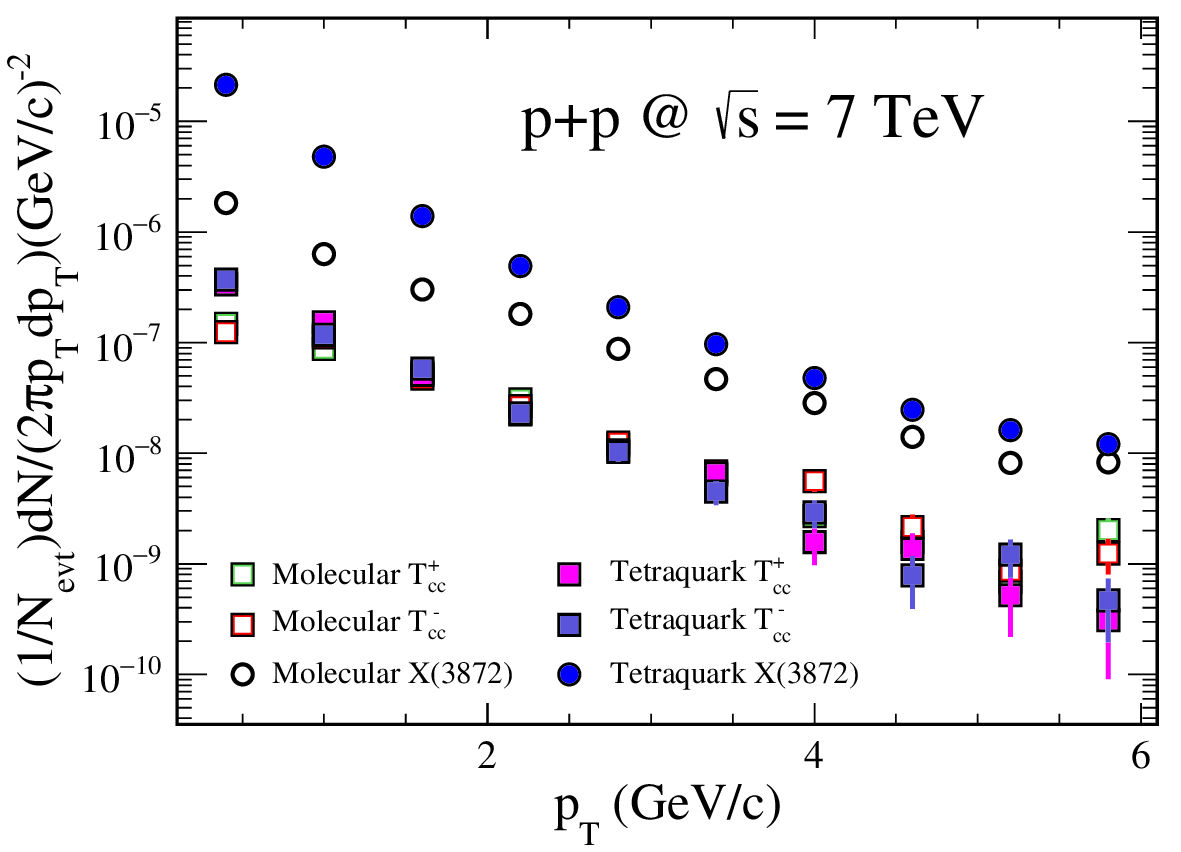}
    \caption{The transverse momentum spectra distribution of $T_{cc}$ and $X(3872)$ within the molecular and tetrquark framework in $pp$ collisions at $\sqrt{s}=7$~TeV. The vertical bars are for statistic uncertainties.}
    \label{spectra}
\end{figure}

To gain further insight, we calculate the transverse momentum spectra of these states under both the molecular and tetraquark states. The distributions are presented in Fig.~\ref{spectra}.
As illustrated in~\cref{spectra}, the $p_T$ spectra exhibit a steep decline with increasing $p_T$, which is consistent with theoretical expectations. Beyond the difference in total yields, the distributions for the molecular and tetraquark states show distinct characteristics. Consequently, the $p_T$ spectra distribution of them in high statistics measurements at experiments can be able to discriminate its structure in the future.

\begin{figure}[htbp!]
  \centering
    \includegraphics[width=0.45\textwidth]{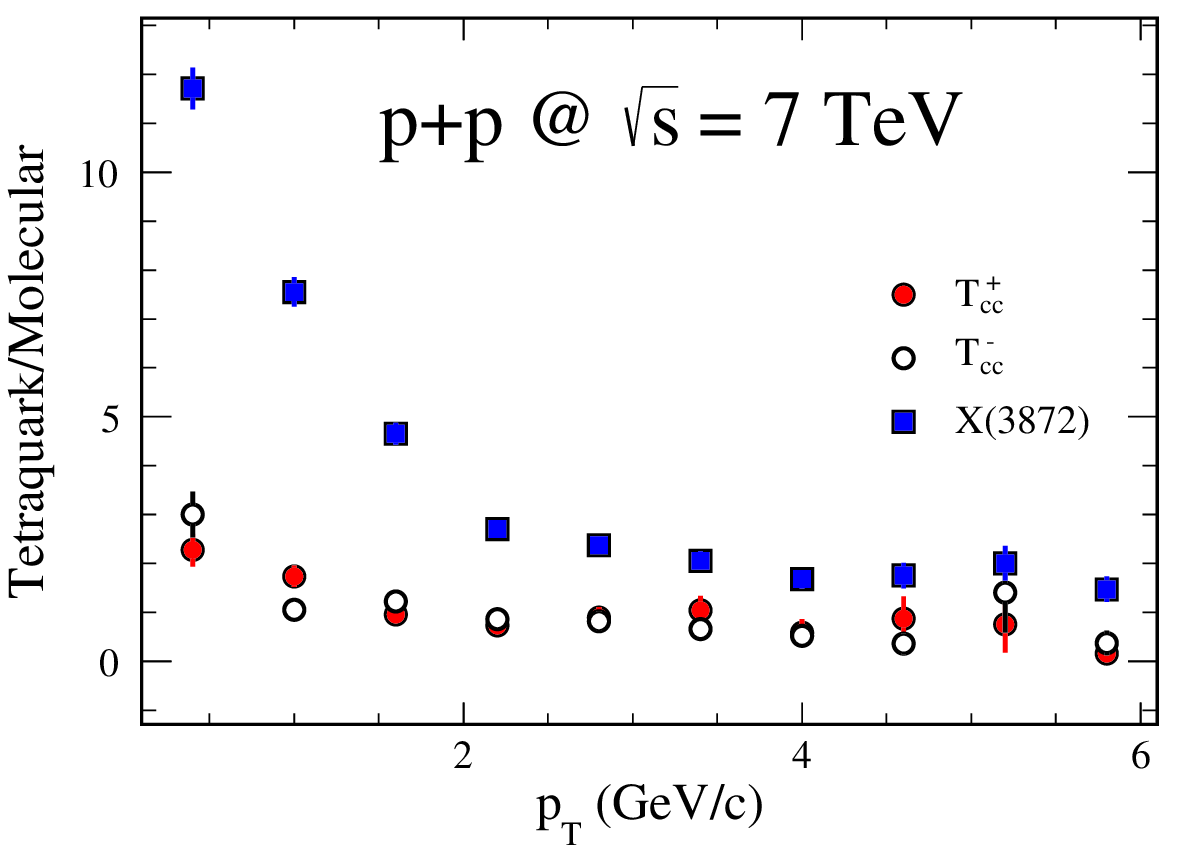}
    \caption{The ratio of yield between the tetraquark and molecular state of $T_{cc}$ and $X(3872)$ in $pp$ collisions at $\sqrt{s}=7$~TeV as functions of $p_T$.}
    \label{ratio}
\end{figure}

To further quantify this difference, we calculate the yield ratios between tetraquark state and molecular state of $T_{cc}^+$, $T_{cc}^-$ and $X(3872)$, respectively, as a function of $p_T$. The distributions of the ratio are shown in Fig.~\ref{ratio}.
The ratio is observed to decrease as $p_T$ increases, characterized by a rapid drop in low $p_T$, accompanied by a more gradual decline in higher $p_T$.
Additionally, the ratio for $X(3872)$ decreases more significantly than that of the $T_{cc}$ states. For $T_{cc}^+$ and $T_{cc}^-$,
the distributions are consistent with each other within error bars. Then, we analyze their asymmetry.

\subsection{Asymmetry}

Ref.~\cite{Hua:2023zpa} concluded that the compact tetraqurak exhibits a significant production asymmetry between $T_{cc}^+$ and $T_{cc}^-$ , enabling the unambiguous determination of the internal structure of the tetraquark.
To discriminate the tetraquark and molecular state of $T_{cc}$, we introduce the asymmetry of the production of $T_{cc}^+$ and $T_{cc}^-$, which is defined as
\begin{equation}
\mathcal{A}\equiv\frac{N_{T_{cc}^-}-N_{T_{cc}^+}}{N_{T_{cc}^-}+N_{T_{cc}^+}}, \label{eq_symmetry}
\end{equation}
where $N_{T_{cc}^+}$ and $N_{T_{cc}^-}$ represent the yield of $T_{cc}^+$ and $T_{cc}^-$, respectively.

\begin{figure}[htbp!]
  \centering
    \includegraphics[width=0.45\textwidth]{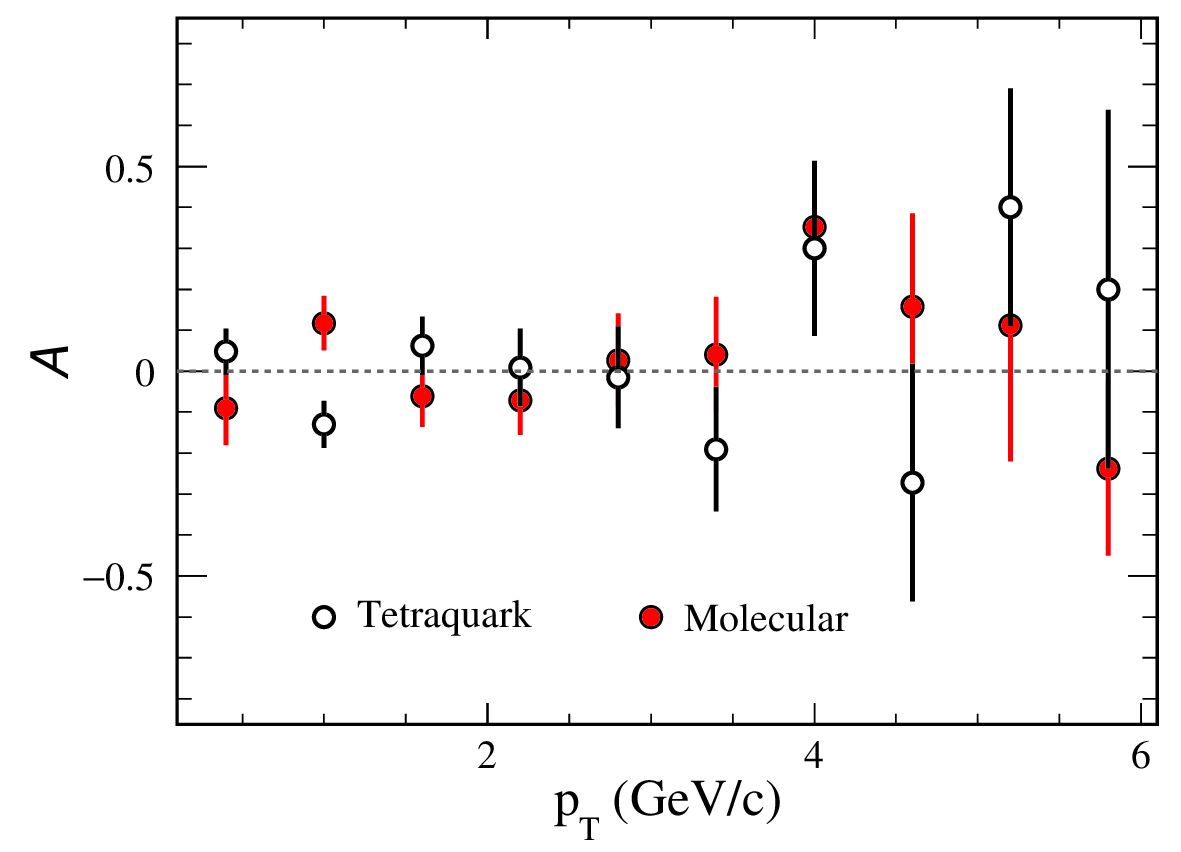}
    \caption{The asymmetry of charged $T_{cc}$ in $pp$ collisions at $\sqrt{s}=7$TeV, for molecular and tetraquark picture, respectively, which defined in Eq.\eqref{eq_symmetry}.}
    \label{symmetry}
\end{figure}

When comparing in two pictures, i.e., molecular and tetraquark picture, the asymmetry is shown in Fig.~\ref{symmetry}.
We can find that  $\mathcal{A}$ presents a smaller production asymmetry in tetraquark state than in molecular state when it is in low $p_T$,
and the asymmetry exhibits a larger production asymmetry in tetraquark states than in molecular states in high $p_T$.
The $\mathcal{A}$ represents a significant discrepancy in two pictures,
therefore, the $\mathcal{A}$ can be a valuable criterion to identify the molecular or tetraquark $T_{cc}$.

\subsection{Coalescence parameter}

\begin{figure*}[htbp!]
  \centering
    \includegraphics[width=0.95\textwidth]{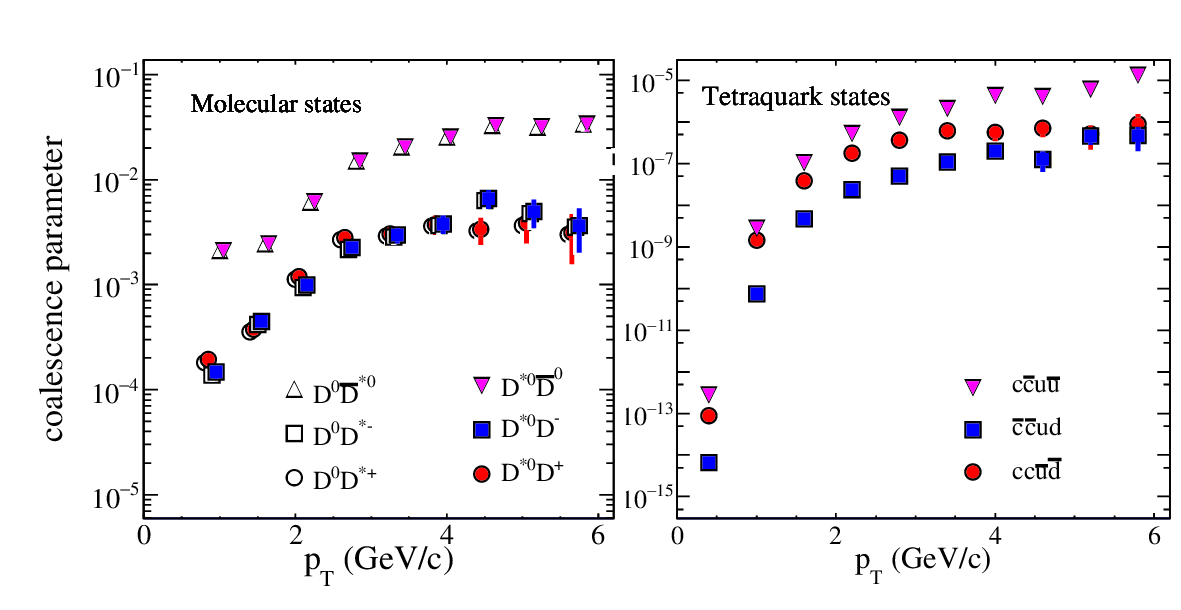} %
    \caption{Coalescence parameter defined in Eq.~\eqref{coalescence_M_2} and \eqref{coalescence_T} of $T_{cc}$ and $X(3872)$ as functions of $p_T$ for molecular(left) and tetraquark(right) picture, respectively, in $pp$ collisions at $\sqrt{s}=7$TeV. The vertical bars represent the statistical uncertainties.}
    \label{coalescence_parmeter}
\end{figure*}

In the coalescence framework~\cite{Butler:1963pp,Scheibl:1998tk}, the production of a cluster is generally regarded as the result of coalescence of constituent particles that are spatially close and have similar velocities at the chemical freeze-out. This framework has been extensively and successfully applied to describe the production of light nuclei and anti-nuclei in $pp$ and heavy-ion collisions~\cite{ALICE:2024yin,STAR:2019sjh}, and we extend it to the study of exotic states $T_{cc}^{\pm}$ and $X(3872)$ in the present work to extract their coalescence parameters. For these exotic states, two distinct structural interpretations (loose molecular state and compact tetraquark state) are considered. The coalescence parameter serves as a robust probe to characterize both the correlation length and correlation volume between constituent particles, as well as the properties of the emission source. Notably, the magnitude of the coalescence parameter is closely related to the correlation volume: a larger coalescence parameter corresponds to a smaller correlation volume, and vice versa. This relation enables us to distinguish whether the exotic states are formed in the hadronic final state (HFS, for molecular states) or the partonic final state (PFS, for tetraquark states) through the measured value of the coalescence parameter.

The invariant yield of a cluster is proportional to the product of the invariant yields of its constituent particles, which is quantitatively described by the general formula:

\begin{equation}
E_X\frac{d^3N_X}{dp^3_X}=B_N\prod_{i=1}^{N}\left(E_i\frac{d^3N_i}{dp^3_i}\right)\label{coalescence_M},
\end{equation}
where $N$ denotes the number of constituent particles of the cluster, $E_X\dfrac{d^3N_X}{dp^3_X}$ is the invariant differential yield of the cluster (e.g., $T_{cc}^{\pm}$, $X(3872)$ in our work), and $E_i\dfrac{d^3N_i}{dp^3_i}$ represents the invariant differential yield of the $i$th constituent particle (quarks for tetraquarks or mesons for molecular states in our work). The coalescence parameter $B_N$ is inversely related to the correlation volume of the emitting source: a larger $B_N$ indicates a more compact source with stronger correlations between constituents.

In general, the production of light nuclei and anti-nuclei can be described by coalescence parameter $B_A$, which has been used extensively and successfully in $pp$~\cite{ALICE:2024yin} and heavy ion~\cite{STAR:2019sjh} collisions.
 %The production of light nuclei and anti-nuclei in $pp$ collisions is expected to be the result of the coalescence of protons and neutrons that are nearby in space and have similar velocities at the last stage of the collision. To quantify this process, the coalescence parameter $B_A$ is introduced, where $A$ represents the mass number of the nucleus under study.
In the coalescence framework, the production of cluster is expected to be the result of the coalescence of the constituents that are nearby in space and have similar velocities at chemical freeze out of mesons or quarks constituents.
To analyze the production of these exotic states , we introduce the coalescence parameters. We consider two distinct structural interpretations: the loose molecular and compact tetraquark state. For the molecular state, the coalescence parameters correspond to $B_2$, which relates the invariant differential yield of exotic states to the one of mesons via the following equation:
\begin{equation}
B_2={(\frac{1}{2\pi}\frac{dN^X}{p_T^Xdp_T^X})}/[{(\frac{1}{2\pi}\frac{dN^D}{p_T^Ddp_T^D})}\cdot {(\frac{1}{2\pi}\frac{dN^{D^*}}{p_T^{D^*}dp_T^{D^*}})}],  \label{coalescence_M_2}
\end{equation}
where the superscript $X$ represents the exotic state($X(3872)$ or $T_{cc}^+$, $T_{cc}^-$ in this work), $D$ and $D^*$ represent the constituent of molecular $X$, the $D$ or $D^*$ meson yield is measured at a value of half of the transverse momentum of exotic state, $p_T^{D/D^*} = \frac{1}{2}p_T^X$.

Similarly, in tetraquark picture, the invariant yield of exotic states is proportional to the invariant yield of quarks(the constituents of the tetraquark state), the coalescence parameter is noted as $B_4$,

\begin{equation}
\begin{aligned}
B_4&={(\frac{1}{2\pi}\frac{dN^X}{p_T^Xdp_T^X})}/[{(\frac{1}{2\pi}\frac{dN^{Q_1}}{p_T^{Q_1}dp_T^{Q_1}})} \cdot {(\frac{1}{2\pi}\frac{dN^{Q_2}}{p_T^{Q_2}dp_T^{Q_2}})} \\
& \cdot {(\frac{1}{2\pi}\frac{dN^{q_1}}{p_T^{q_1}dp_T^{q_1}})} \cdot {(\frac{1}{2\pi}\frac{dN^{q_2} }{p_T^{q_2}dp_T^{q_2}})}],  \label{coalescence_T}
  \end{aligned}
\end{equation}
where the superscript $X$ represents the exotic state, $Q$ and $q$ represent the quark constituents of $X$ ($Q=Q_1=Q_2=c/\bar{c},q=q_1=q_2=u/\bar{u}/d/\bar{d}$), the yield of quarks is calculated at a value of transverse momentum commensurate their mass, $p_T^{Q/q} = \frac{m^{Q/q}}{M}p_T^X$, ($M=m^{Q_1}+m^{Q_2}+m^{q_1}+m^{q_2}$).
%the yield of quarks at a value of quart of transverse momentum of the exotic state, $p_T^{Q/q} = \frac{1}{4}p_T^X$(the mass of quarks has been taken into account).

Based on $p_T$ of exotic states and their constituents, the coalescence parameter $B_2$ and $B_4$ are calculated 
as functions of $p_T$ and shown in Fig.~\ref{coalescence_parmeter}, respectively.
Our results show a rising trend in the coalescence parameters with increasing $p_T$. Since a larger coalescence parameter corresponds to a more compact source,
this trend implies that the correlation volume decreases with increasing $p_T$.
The observation is consistent with the principle that the coalescence is inversely proportional to the correlation volume of the emitting source.

\section{summary and outlook}\label{sec4}
In this work, we simulated the production of $T_{cc}^+$, $T_{cc}^-$ and $X(3872)$ states within both molecular and tetraquark frameworks in $pp$ collision at $\sqrt{s}=$ 7 TeV. Our comparison reveals that the yields of $T_{cc}$ are consistently lower than those of $X(3872)$, which may be attributed to the ``threshold" effect of the required double charm quarks for $T_{cc}$ formation. Furthermore, the yields of the tetraquark states are found to be significantly larger than those for the molecular states. The analysis of the $p_T$ spectral distribution shows distinct differences between the molecular and tetraquark state, suggesting that transverse momentum spectra can serve as an effective tool to discriminate between these internal structures in future high-statistics experiments.
The production asymmetry $\mathcal{A}$ exhibits distinctly different behaviors between the molecular and tetraquark scenarios for $T_{cc}^+$ and $T_{cc}^-$
in different $p_T$ regions.
We also find that the coalescence parameter increases monotonically as a function of $p_T$. Interestingly, as shown in Fig.~\ref{coalescence_parmeter}, no differences are found in the distribution shapes of a given state between different constituents.

Given the advantage of heavy ion collisions in producing an abundance of the charmed and doubly charmed exotic states, our future research will extend to this environment to explore the production and properties of these states.

\section{Acknowledgements}
This work is supported by the Scientific Research Foundation of Research Project of Hubei Provincial Department of Education (No. D20233003 and No. B2023191).

\bibliography{ref}% Produces the bibliography via BibTeX.

\end{document}